\documentclass[a4paper,mathleft]{an}
\usepackage{graphicx}
\usepackage{epsfig}
\usepackage{times}
\begin{document}


\title{A framework for describing correlated excitation of solar p
modes}

\author{W.~J.~Chaplin\fnmsep\thanks{Corresponding author:
\email{w.j.chaplin@bham.ac.uk}\newline} \and Y.~Elsworth \and
T.~Toutain}

\authorrunning{W. J. Chaplin et al.}

\institute{School of Physics and Astronomy, University of Birmingham,
Edgbaston, Birmingham, B15 2TT, UK}

\received{20 Oct 2007}
\accepted{XX Xxx 200X}
\publonline{later}


\abstract{In a previous paper we suggested that, for a given p mode,
the excitation function is the same as the component of the solar
background noise that has an identical surface spherical harmonic
projection (over the corresponding range of temporal frequency). An
important consequence of this surmise is that the excitation of
overtones of a given angular degree and azimuthal order will be
correlated in time. In this note, we introduce the basic principles
and a mathematical description of correlated mode excitation. We use
simple, illustrative examples, involving two modes.  Our treatment
suggests that in the real observations, any signatures of the
correlation would not appear as a correlation of the output amplitudes
of overtones, but rather as subtle modifications to the power spectral
density at frequencies between the central frequencies of the
overtones. These modifications give a contribution to the observed
peak asymmetries.}

\maketitle

 \section{Introduction}
 \label{sec:intro}

It has now been well established that solar p-mode peaks observed in
the frequency power spectrum have profiles that are asymmetric in
frequency. The modes are excited stochastically in a thin layer at top
of the convection zone, and the asymmetry is believed to result from
important characteristics of this excitation (e.g., see Nigam \&
Kosovichev 1998; Nigam et al. 1998; Kumar \& Basu 1999a, b; Severino
et al. 2001; Jefferies et al. 2003). First, there is a contribution to
the asymmetry from the spatial location, radial extent, and multipole
properties of the acoustic source. Second, there is a contribution
from what is commonly referred to as \emph{correlated background
noise}. This background noise is signal from the convective
granulation. Since the convection gives rise to the acoustic source
that excites the modes, the granulation and modes are correlated.
These correlations give rise to asymmetry. It is the observational
consequences of such correlations that are the subject of this paper.

In Toutain, Elsworth \& Chaplin (2006), we presented a framework to
explain the contribution of the correlated noise to the asymmetry. We
hypothesized that the excitation function of a mode of angular degree
$l$, azimuthal degree $m$, and frequency $\nu_0$, is the same as that
component of the solar background (granulation) noise that has the
same spherical harmonic projection, $Y_{lm}$, in the corresponding
range in frequency in the Fourier domain. It is this component of the
noise background that gives a contribution to the asymmetry.

An important implication of the above is that overtones of a given
($l$,$m$) should have excitation functions that are correlated in
time. Indeed, correlation of the excitation follows naturally from
invoking correlations with the background noise. Correlations are not
expected between different components of the same mode. That is
because the $Y_{lm}$ for ($l$,$m$) and ($l$,$-m$) are orthogonal, and
are therefore assumed to have independent, i.e., uncorrelated,
excitation.

Our main aim in this paper is to introduce the basic principles and a
mathematical description of correlated p-mode excitation. The paper is
didactic in nature, and uses simplified examples -- involving just two
modes -- to illustrate the principles.  The paper should be viewed as
providing the necessary background, and framework, needed to consider
the more complicated case of the full spectrum of overtones of a
Sun-like oscillator. We leave a description of how the effects are
manifested in the solar p-mode spectrum to an upcoming paper.

It is important to stress that correlation of the excitation in time
does \emph{not} imply correlation of the mode amplitudes in time. This
can be understood by considering the analogy of damped, stochastically
forced oscillators. Modes of different frequencies will be `kicked' by
the common excitation at different phases in their oscillation cycles,
and provided the frequency differences are at least several times
greater than the peak linewidths, there will be significant
differences in how the amplitudes vary in time due to the excitation,
a point we return to in Section~\ref{sec:corramp} below. The
correlation is instead manifested in a more subtle manner in the
frequency power spectrum: specifically, it modifies the power spectral
density of the spectrum at frequencies between the central frequencies
of the overtones, where the wings of the Lorentzian-like mode profiles
interact, and gives a contribution to the observed peak asymmetry. It
is tempting to say this another way: that ``the background is
modified'' between the modes. However, that would be a misleading
statement, in that with correlated background and correlated
excitation the clear distinction between modes and background no
longer exists. The next generation of `peak bagging' mode-fitting
codes will need to account for this blurred distinction.

The layout of the rest of the paper is as follows. We begin in
Section~\ref{sec:corramp} by showing that if two modes are excited by
the same functions in time, this does not always imply the oscillation
amplitudes will also be correlated in time. We then turn to the issue
of the mode profile asymmetry, beginning in Section~\ref{sec:recap}
with a recap of the discussion in Toutain, Elsworth \& Chaplin (2006)
on how correlated background noise gives rise to mode peak asymmetry.
This is followed by the main part of our paper: In
Section~\ref{sec:two} we show how correlation between the excitation
of different modes modifies the observed power spectral density
(analytical descriptions are provided in Appendix~\ref{sec:app}). We
finish in Section~\ref{sec:disc} by discussing the implications of
these modifications, flagging work in progress.

 \section{Time correlation of amplitudes of different modes}
 \label{sec:corramp}

In order to illustrate the impact of correlated mode excitation on
correlations in time between the output amplitudes of different modes,
we made simulations of stochastic harmonic oscillator timeseries. The
Laplace transform solution of the equation of a forced, damped
harmonic oscillator was used to generate timeseries of the output
velocity of artificial modes on a 40-sec cadence, in the manner
described by Chaplin et al. (1997).  The oscillators were excited at
each time sample with small `kicks' from timeseries of random Gaussian
noise.

For each simulation we generated artificial timeseries realizations of
two modes. Both modes were excited by the same timeseries of kicks,
meaning their excitation was 100\,\% correlated in time. The first
mode always had a natural frequency of $3010\,\rm \mu Hz$; the
frequency of the second mode was changed from one simulation to
another, and, over the full sequence of simulations covered the range
2990 to $3010\,\rm \mu Hz$. We chose this range to illustrate results
for two correlated modes, closely spaced in frequency.  Both modes
always had intrinsic damping rates that gave natural linewidths of
$\Gamma=1\,\rm \mu Hz$ for their resonant peaks. This linewidth
corresponds approximately to the linewidth of modes at the centre of
the low-$l$ solar p-mode spectrum, and is equivalent to a lifetime,
$\tau = 1/ \pi \Gamma$, of $\sim 3.7\,\rm days$. We comment briefly
below on how the results are affected by the choice of $\Gamma$.

For the analysis, we used 0.5-day-long timeseries. This meant the mode
peaks were unresolved in the frequency domain. From each pair of
0.5-day timeseries, we computed the RMS velocity amplitudes of the
first and second modes. We repeated each simulation 100 times, using a
new timeseries of random Gaussian noise to excite both modes. We then
computed the Pearson correlation coefficient between the 100 measured
RMS velocity amplitudes of the first and second modes. This gave us a
single correlation measure for a given pair of input frequencies. We
then made new sets of 100 simulations for different input frequencies
of the second mode.


\begin{figure}
 \centerline {\epsfxsize=7.0cm\epsfbox{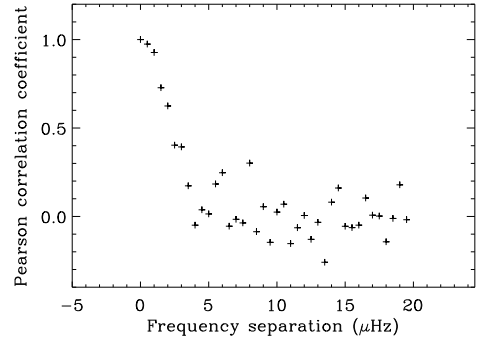}}

 \caption{Correlation of the output amplitudes of two modes, as
 measured from 0.5-day timeseries, as a function of the separation in
 frequency between the two modes. The excitation of both modes is
 always 100\,\% correlated in time.}

 \label{fig:corrslip}
\end{figure}


A plot of the measured correlations of the output amplitudes is
presented in Figure~\ref{fig:corrslip}. The correlations are plotted
as a function of the separation in frequency, $\Delta\nu$, between the
two modes. When the frequency separation was zero, both modes had a
frequency of $3010\,\rm \mu Hz$. When the separation was $20\,\rm \mu
Hz$, the first mode had a frequency of $3010\,\rm \mu Hz$ and the
second mode had a frequency of $2990\,\rm \mu Hz$. It is readily
apparent that when the frequencies differ by more than $\approx 5\,\rm
\mu Hz$, the Pearson correlation coefficient is scattered around
zero. Only at small frequency separations does the correlation of the
output amplitudes become significant (and of course reaches unity for
zero separation). The rate at which the correlation falls with
increasing separation of the input frequencies depends on the ratio
$\Delta\nu / \Gamma$. Figure~\ref{fig:corrslip} shows that the
correlation coefficient drops to 0.5 when $\Delta\nu / \Gamma \approx
2$. So an increase of $\Gamma$ will result in the coefficient of
correlation falling off more slowly (and vice versa).

Consecutive overtones of the low-$l$ solar p modes are actually
separated in frequency by $\Delta\nu \sim 135\,\rm \mu Hz$, which is
the so-called large frequency spacing. This separation is almost seven
times larger than the maximum separation shown in
Figure~\ref{fig:corrslip}.  If our hypothesis regarding the correlated
mode excitation is correct -- this being that it is the excitation of
overtones of a given ($l$,$m$) that is correlated -- the results here
indicate the measured amplitudes of the modes will not be correlated
in time.

 \section{Recap: impact of correlated background noise}
 \label{sec:recap}

Consider a p mode that is stochastically excited and intrinsically
damped. The excitation function is assumed to be a random function
with zero mean and unit variance (the strength of the excitation is
folded into the mode amplitude).  The excitation function in the
frequency domain, $E(\nu)$, may therefore be written as:
 \begin{equation}
 E(\nu) = e_r(\nu) + i e_i(\nu).
 \end{equation}
Similarly, the noise is assumed to be a random function with unit
variance, multiplied by a frequency-dependent amplitude, i.e.,
 \begin{equation} 
 N(\nu) = \sqrt{\frac{n(\nu)}{2}} [n_r(\nu) + i
 n_i(\nu)].  
 \label{eq:N}
 \end{equation}
The frequency response of the mode, which we assume may be modelled as
a forced, damped oscillator of high Q, is described as a Lorentzian,
i.e.,
\begin{equation} L(\nu) = \frac{x}{1+x^2}
\sqrt{H/2} + i \frac{1}{1+x^2}\sqrt{H/2}, \label{eq:lor}
\end{equation}
where $H$ is the maximum power spectral density (mode `height'),
$x=(\nu-\nu_{0})/(\Gamma/2$), and $\nu_0$ and $\Gamma$ are the central
frequency and linewidth of the mode, respectively.  Following, for
example, the discussion in Anderson et al. (1990), we relate these
quantities to determine the solution, in the Fourier (frequency)
domain, of the oscillator equation describing the behaviour of the p
mode:
 \begin{equation}
 V(\nu) = L(\nu) E(\nu) + N(\nu).
 \end{equation}
If we assume that the excitation function and the background noise are
uncorrelated the limit power spectral density is described by:
 \begin{equation} P(\nu) = \left<|V(\nu)|^2\right> =
 \frac{H}{1+x^2} + n(\nu),
 \end{equation}
which includes the usual Lorentzian profile. Here, the angled brackets
($\left< ... \right>$) indicate an average over a large number of
realizations of the excitation function, and of the background noise.

Let us now assume that the excitation function and the background
noise of the observed spectrum are correlated, so that in each
frequency bin
 \begin{equation}
 <e_r n_r> = <e_i n_i> = \alpha,
 \label{eq:alpha}
 \end{equation}
where $\alpha$ is therefore a coefficient of correlation between the
background noise and the excitation function.  The limit power
spectral density then takes the more complicated form:
 \begin{equation}
 P(\nu) = \frac{H}{1+x^2} \left( 1 + 2 \alpha \sqrt{n/H} x \right) + n(\nu).
 \label{eq:ass}
 \end{equation}
For non-zero $\alpha$, this gives an asymmetric profile.

\begin{figure*}
 \centerline 
{\epsfxsize=8.0cm\epsfbox{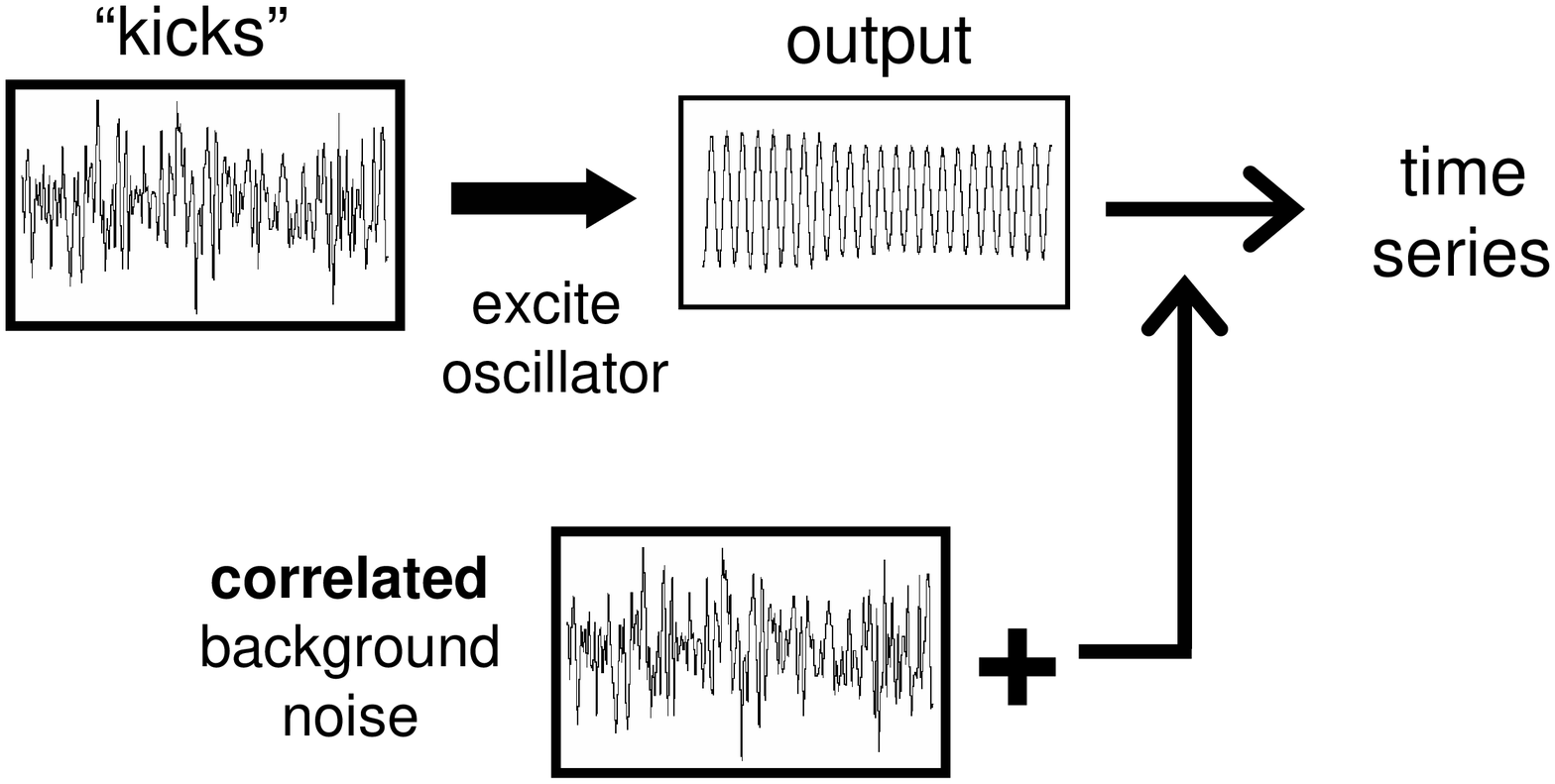}
 \epsfxsize=7.0cm\epsfbox{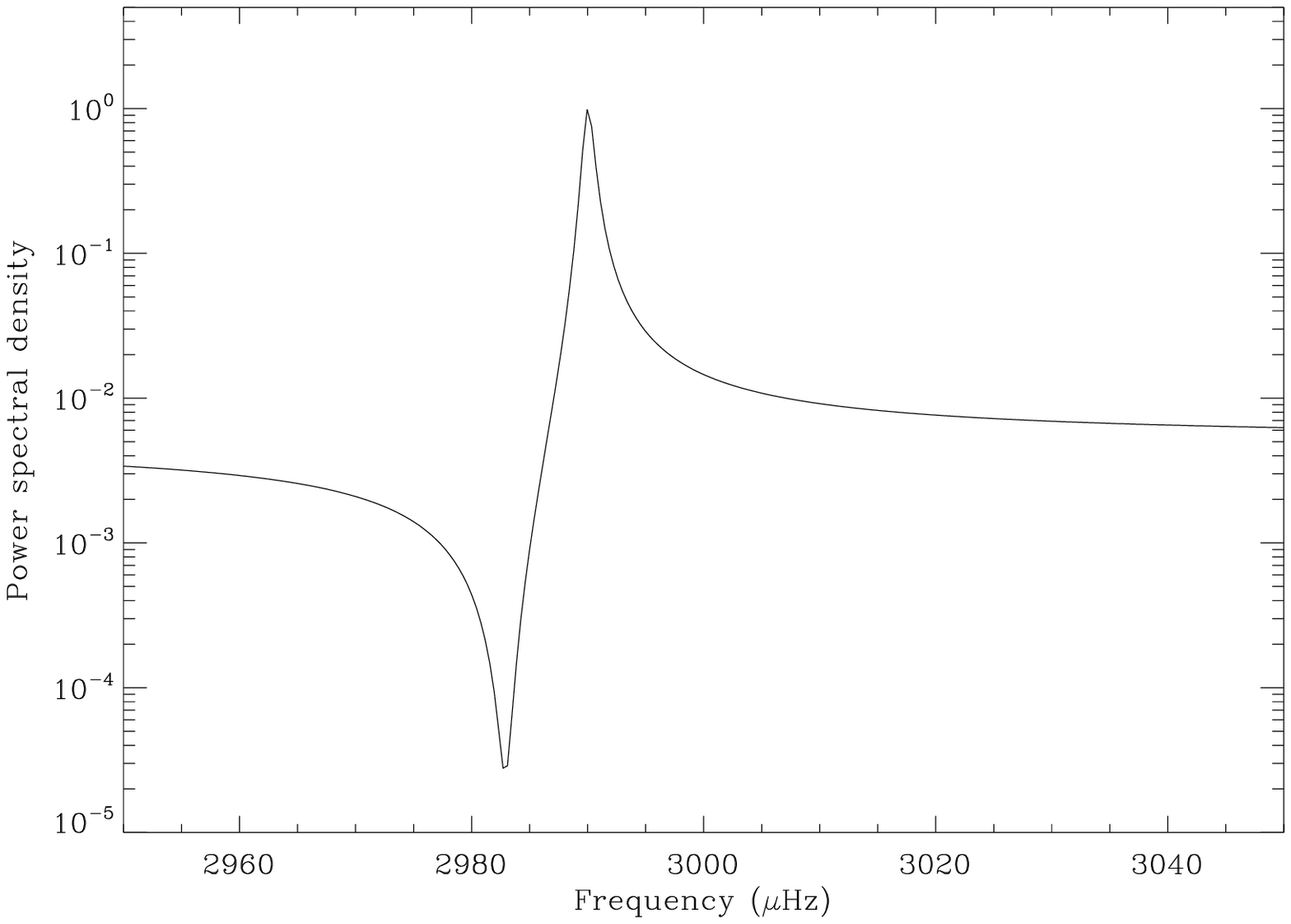}}

 \caption{Case\,\#1: Excitation of a single mode. Correlated
 background noise is added, which is 100\,\% correlated with the
 excitation function.}

 \label{fig:corr0}
\end{figure*}

This scenario is illustrated schematically in Figure~\ref{fig:corr0},
which we call Case\,\#1. The excitation function is represented by a
timeseries of `kicks' of random Gaussian noise.  The kicks excite an
oscillator. In the example shown, the oscillator has a natural
frequency of $\nu_0=2990\,\rm \mu Hz$ and a linewidth of
$\Gamma=1\,\rm \mu Hz$. When the timeseries of kicks is added to the
output of the oscillator, and is therefore used as background, we will
have \emph{correlated} background noise. A direct consequence is that
the frequency power spectrum of the final timeseries will show a peak
with asymmetry.  The underlying, or limit, frequency power spectrum
for Case\,\#1 is shown in the right-hand panel of
Figure~\ref{fig:corr0}. 

In the example shown we scaled the timeseries of kicks to give a
background-to-height ratio of $n/H=0.5\,\%$. We did not add any
uncorrelated background to the timeseries, so that the background
noise was 100\,\% correlated with the excitation (i.e.,
$\alpha=1$). Because this correlation was positive, the sign of the
peak asymmetry was also positive (i.e., more power on the
high-frequency side of the resonance). Negative correlation will give
a peak showing negative asymmetry.


Finally in this section, we note that for Case\,\#1 -- and for the
examples that follow in Section~\ref{sec:two} below -- the frequency
spectrum of the excitation was white, because the timeseries of kicks
had a Gaussian distribution in time. This is a reasonable, first-order
approximation to the solar case, since locally the spectrum of the
solar granulation is white. One could of course be more sophisticated:
In the latest version of the solarFLAG helioseismology simulator,
artificial p modes are excited with random noise having a more
realistic granulation-like frequency spectrum, which has power that
rises with decreasing frequency (Chaplin et al., in preparation; see
also Chaplin et al. 2006, for information on solarFLAG).

 \section{Impact of correlated excitation of different modes}
 \label{sec:two}

We have seen that when the excitation function of a mode is correlated
with the background noise, the profile of the mode is asymmetric. Now,
we ask what happens to the observed power spectral density when there
is correlation of the excitation of different modes. We take the
simple case of a frequency power spectrum that is comprised of two
modes.

As noted previously, consecutive overtones of the low-$l$ solar p
modes are separated in frequency by $\sim 135\,\rm \mu Hz$. Here, we
consider two modes separated in frequency by $20\,\rm \mu Hz$. This
smaller frequency spacing exaggerates the impact of the correlated
excitation on the observed power spectral density, and therefore
allows us to show more clearly the effect of the correlation. We
return again briefly to this point in Section~\ref{sec:disc} below. We
only consider the case of 100\,\% correlation, again, as an
illustrative example of the effects.  In the real solar p-mode
spectrum differences in, for example, the radial dependence of the
excitation with frequency might be expected to reduce the
correlations. The impact of changing the magnitude of the correlations
is considered, in the context of the full solar p-mode spectrum, in a
future paper (Chaplin et al., in preparation).

\begin{figure*}
 \centerline 
{\epsfxsize=8.0cm\epsfbox{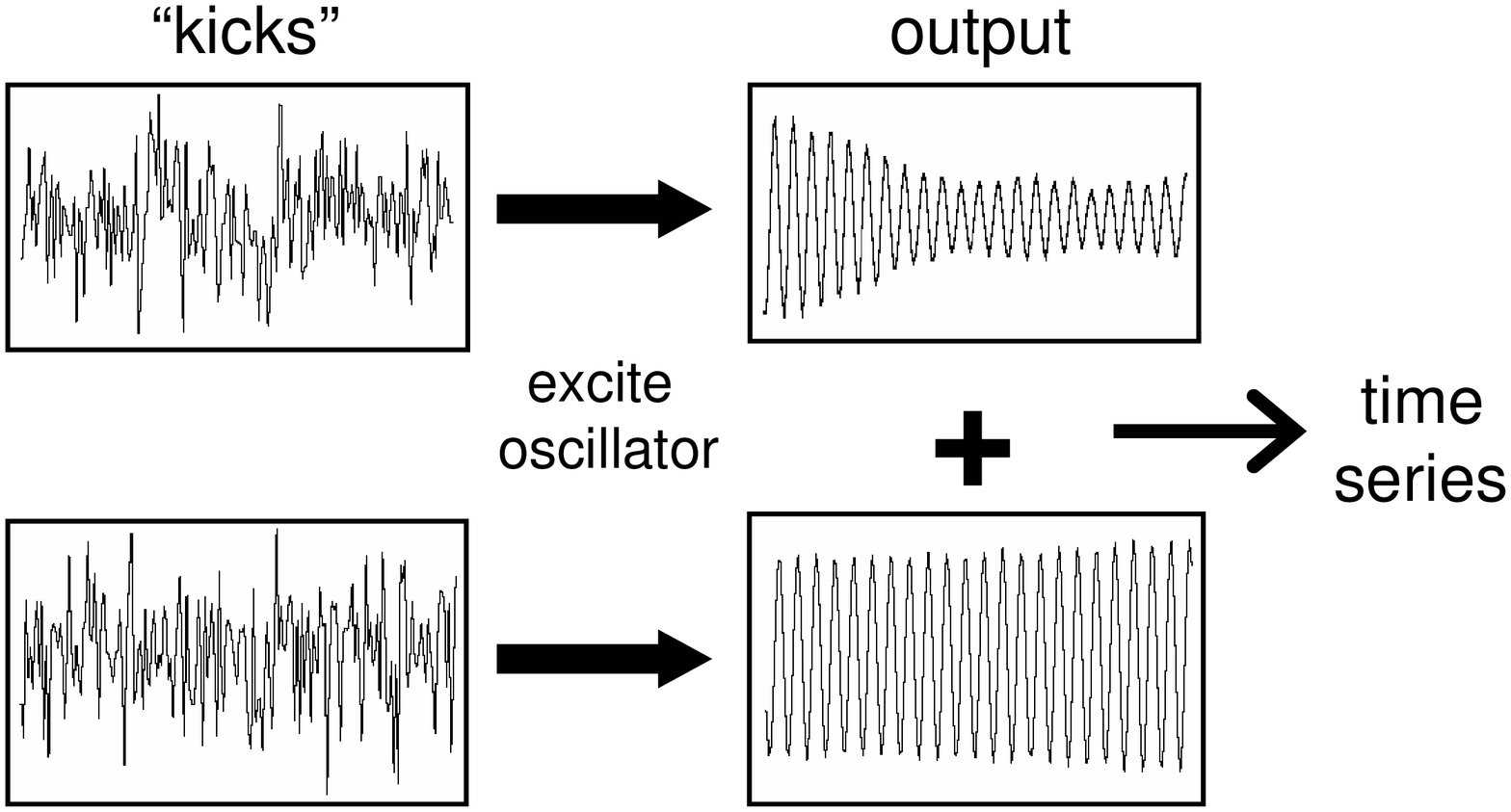}
 \epsfxsize=7.0cm\epsfbox{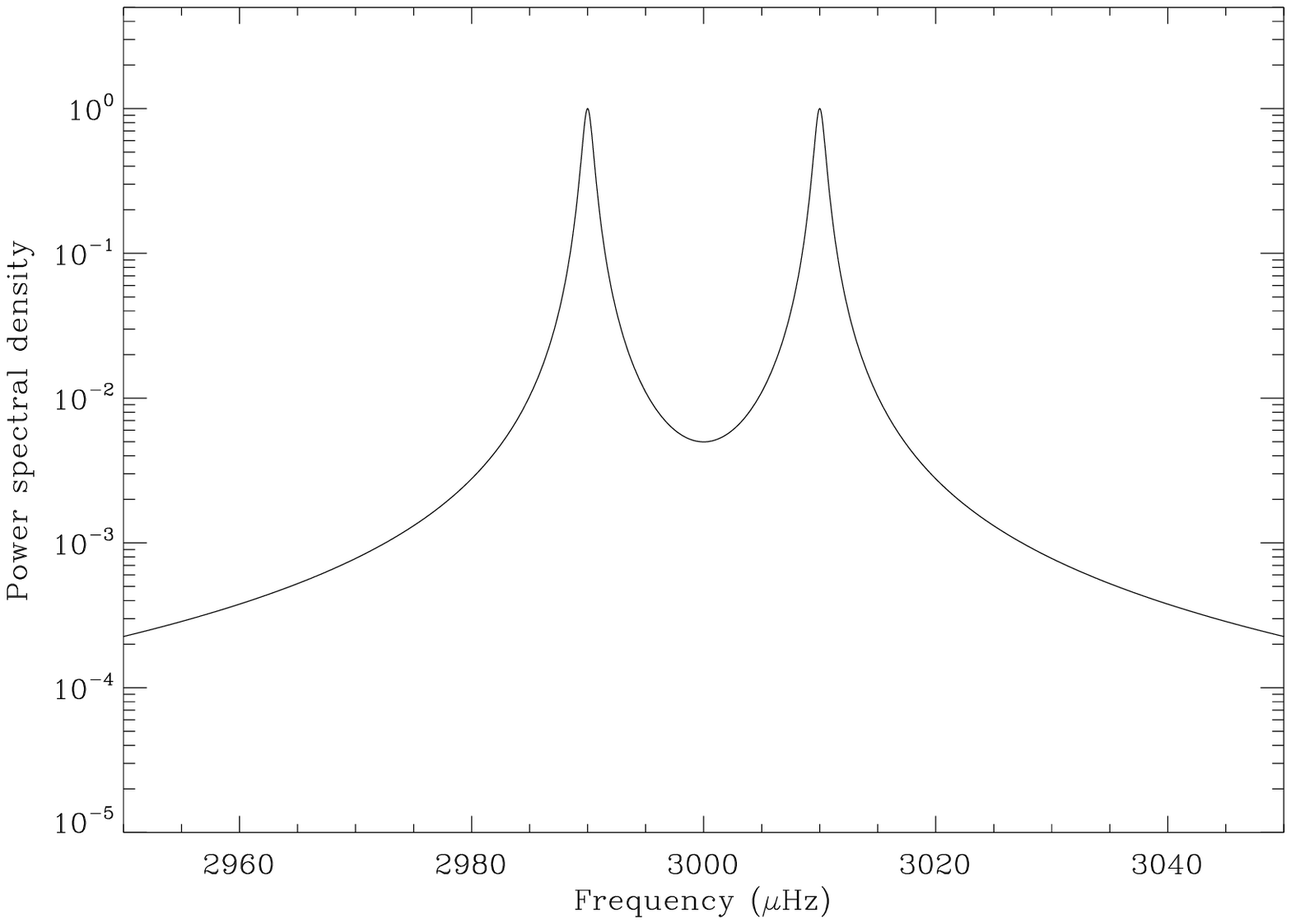}}

 \caption{Case\,\#2: The excitation of two modes is independent, and
 no background noise is added.}

 \label{fig:corr1}
\end{figure*}

We begin with a simple reference case, which we call Case\,\#2. We
again assume each mode has a linewidth of $1\,\rm \mu Hz$. The
excitation function of each mode is assumed to be independent in
time. Furthermore, we do not include any background noise in the
timeseries. The key elements of Case\,\#2 are shown schematically in
the left-hand panel of Figure~\ref{fig:corr1}. The timeseries of
`kicks' excite the oscillators (modes) giving the outputs of velocity
as a function of time, which are then summed to give the final
timeseries. The underlying frequency power spectrum of the final
timeseries is shown in the right-hand panel of
Figure~\ref{fig:corr1}. Since the excitation is independent, the
observed frequency power spectrum for Case\,\#2 is given by the
(incoherent) addition of the frequency power spectra of the
independent oscillator outputs.


\begin{figure*}
 \centerline 
{\epsfxsize=8.0cm\epsfbox{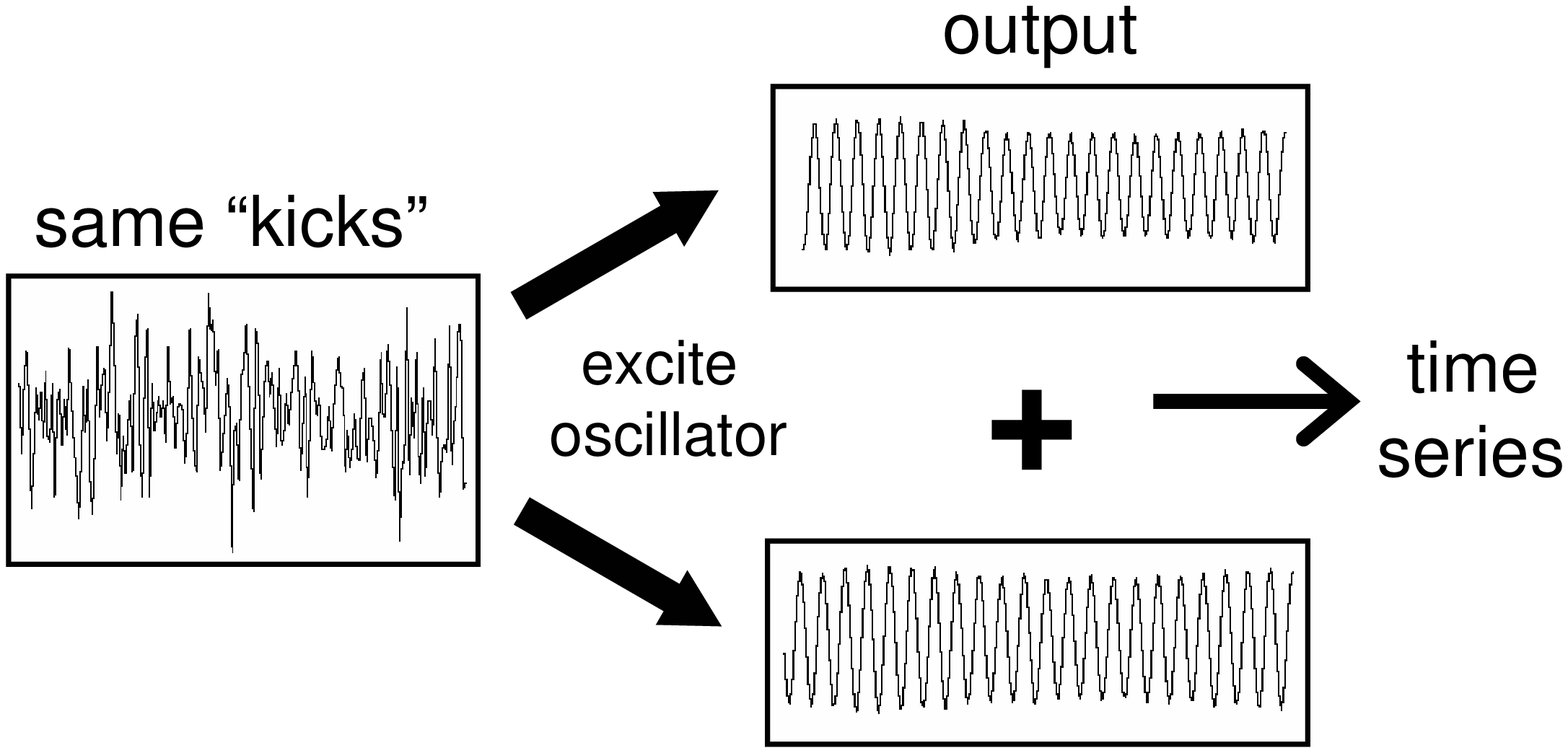}
 \epsfxsize=7.0cm\epsfbox{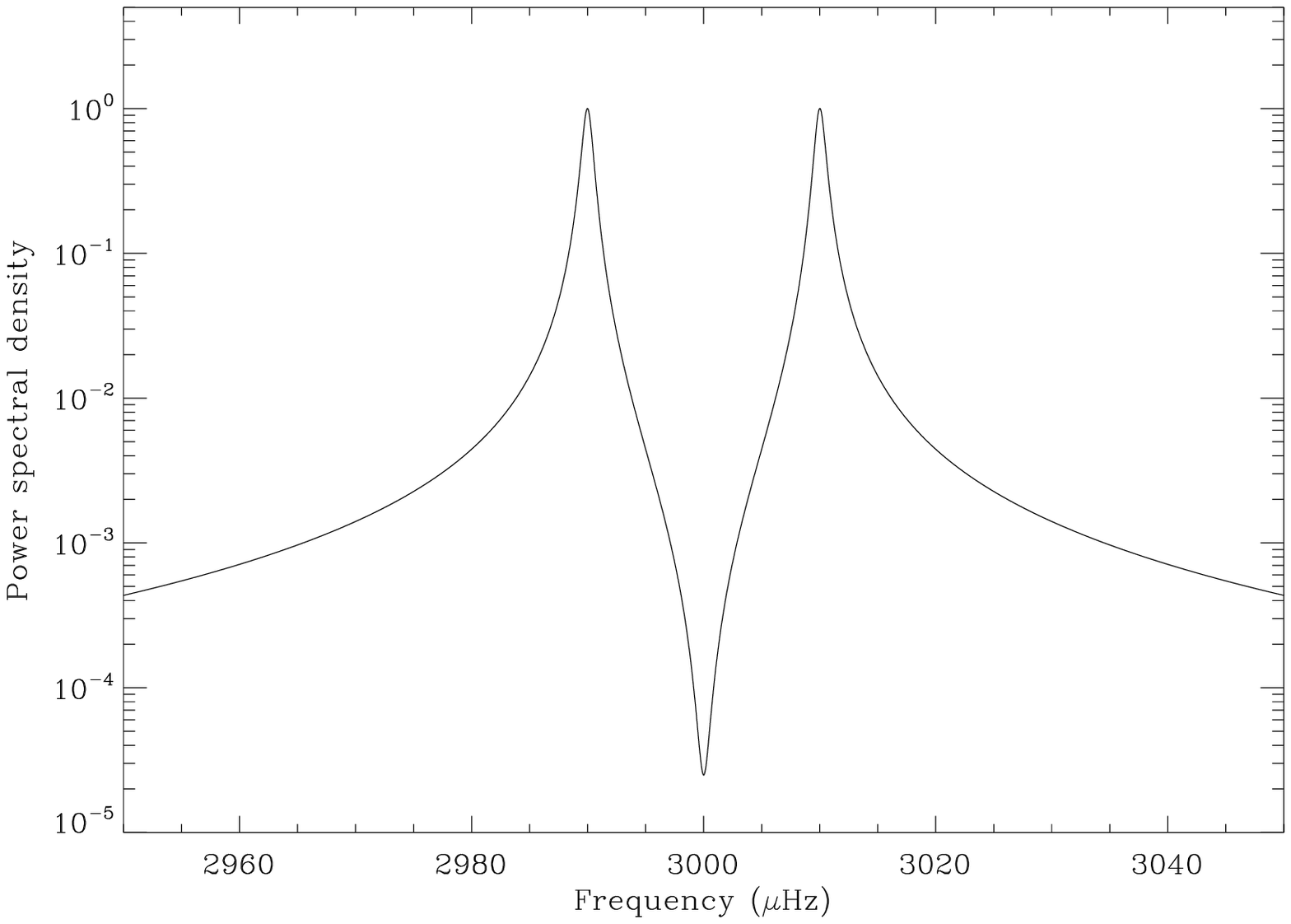}}

 \caption{Case\,\#3: The excitation of two modes is 100\,\%
 correlated, but no background noise is added.}

 \label{fig:corr2}
\end{figure*}

Next, for Case\,\#3, we assume the excitation of the two modes is
100\,\% correlated in time. This case is shown in the left-hand panel
of Figure~\ref{fig:corr2}, where the same timeseries of random
Gaussian noise is used to excite both oscillators. Again, we include
no background noise in the timeseries. The frequency power spectrum of
the final timeseries shows clearly that the peaks are asymmetric. We
stress that this asymmetry does not come from the influence of
correlated background noise (e.g., as in the single-mode case
illustrated in Figure~\ref{fig:corr0}): no background noise was
included here. Instead, it comes from the interaction of the two
modes, whose excitation is correlated. An analytical description of
the power spectral density is presented in Appendix~\ref{sec:case3}.

We draw an important conclusion from Case\,\#3: correlated mode
excitation can give a contribution to the observed asymmetry of
modes.


\begin{figure*}
 \centerline 
{\epsfxsize=8.0cm\epsfbox{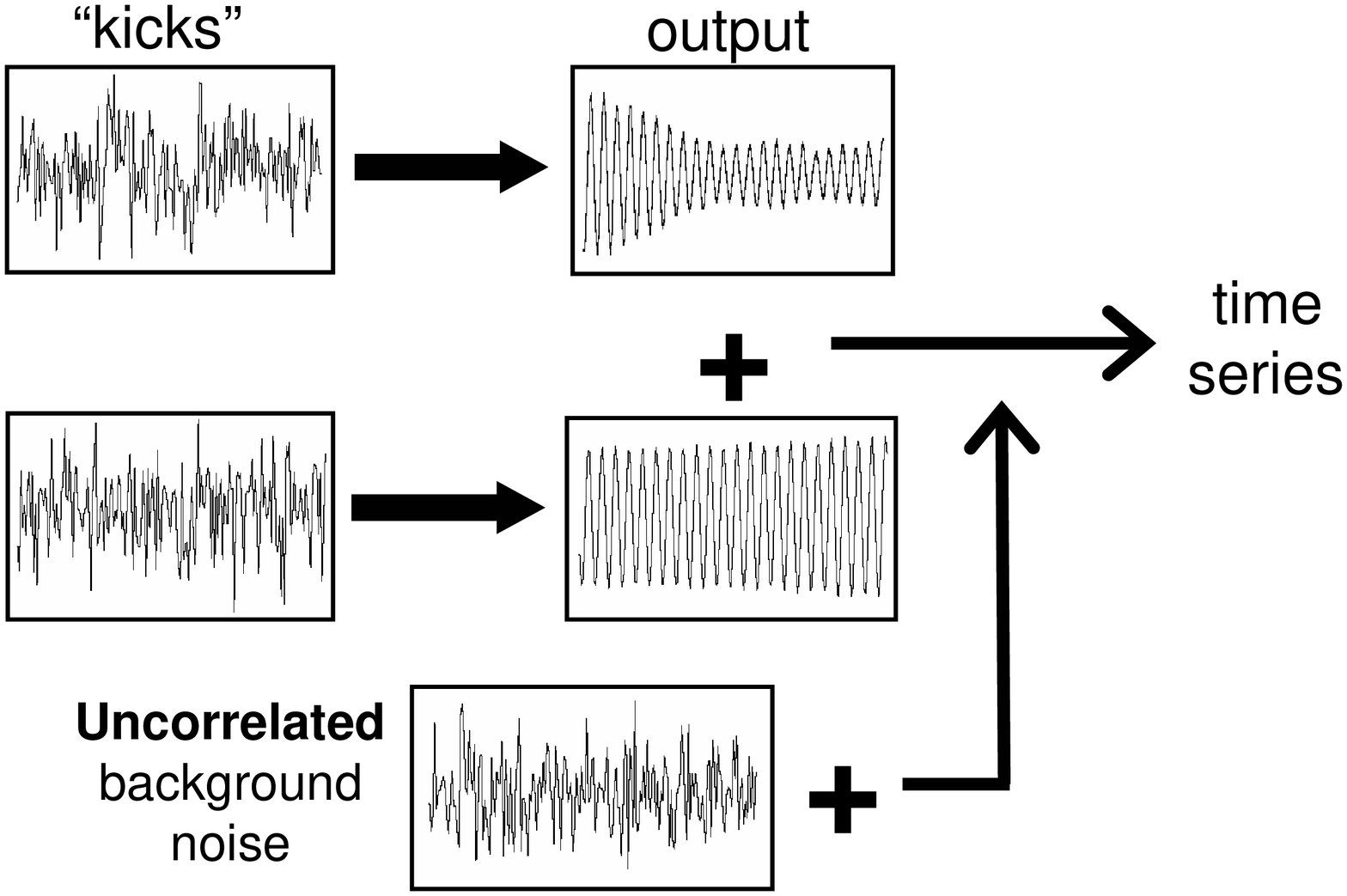}
 \epsfxsize=7.0cm\epsfbox{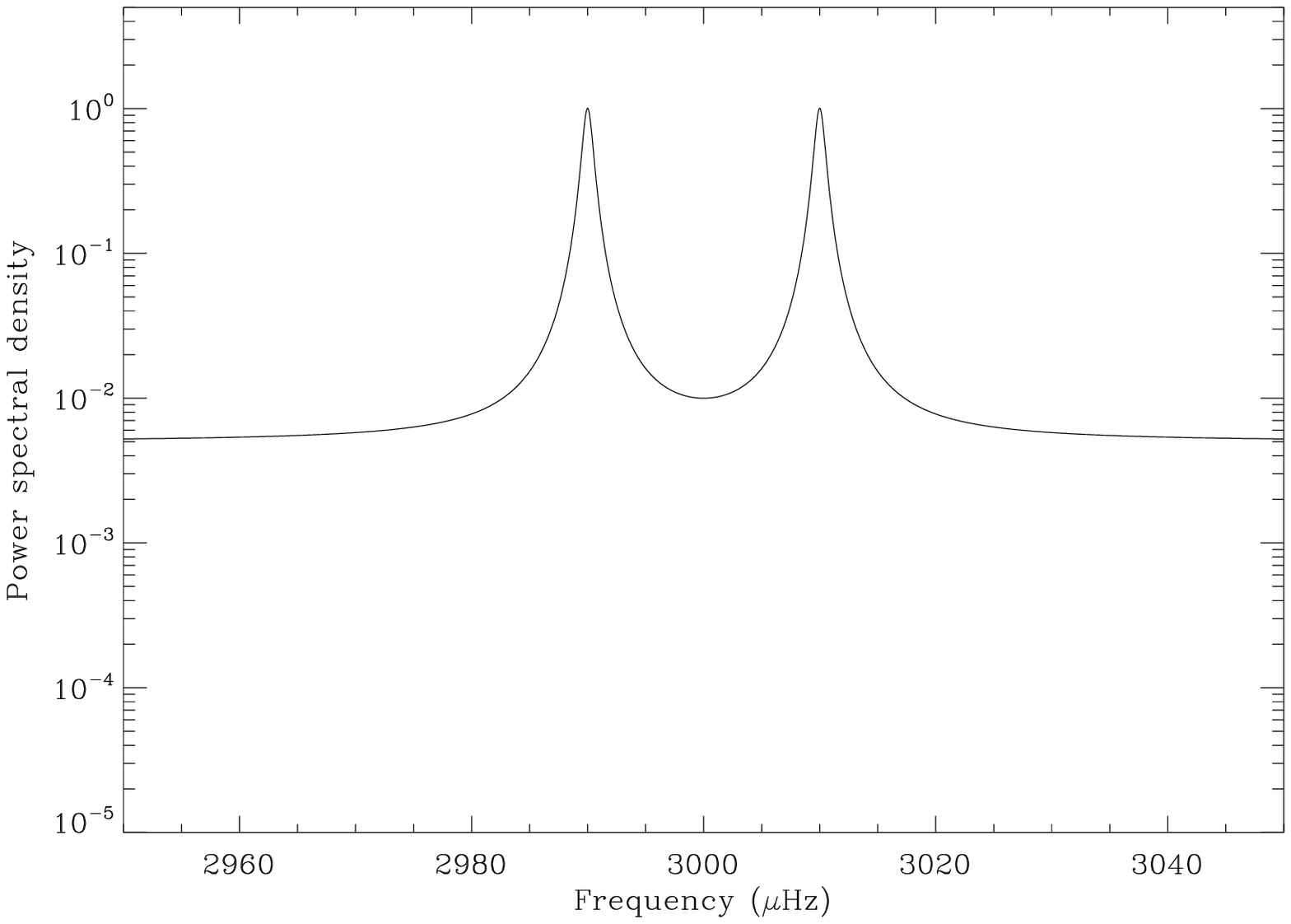}}

 \caption{Case\,\#4: The excitation of two modes is
 independent. Unlike Case\,\#2 the timeseries also includes
 uncorrelated background noise.}

 \label{fig:corr3}
\end{figure*}

Real frequency power spectra include contributions from
background. Let us therefore consider what happens when we add
background noise to the timeseries shown in Case\,\#2 and Case\,\#3
above. For Case\,\#4 (Figure~\ref{fig:corr3}), we add some
independent, uncorrelated Gaussian noise to the final timeseries of
Case\,\#2. The addition of this background noise to the final
timeseries gives rise to a constant background in the frequency power
spectrum. Since the excitation functions and the background are all
uncorrelated, the observed power spectral density is given by the
incoherent addition of the frequency power spectra of each oscillator
in turn, and the background noise.


\begin{figure*}
 \centerline 
{\epsfxsize=8.0cm\epsfbox{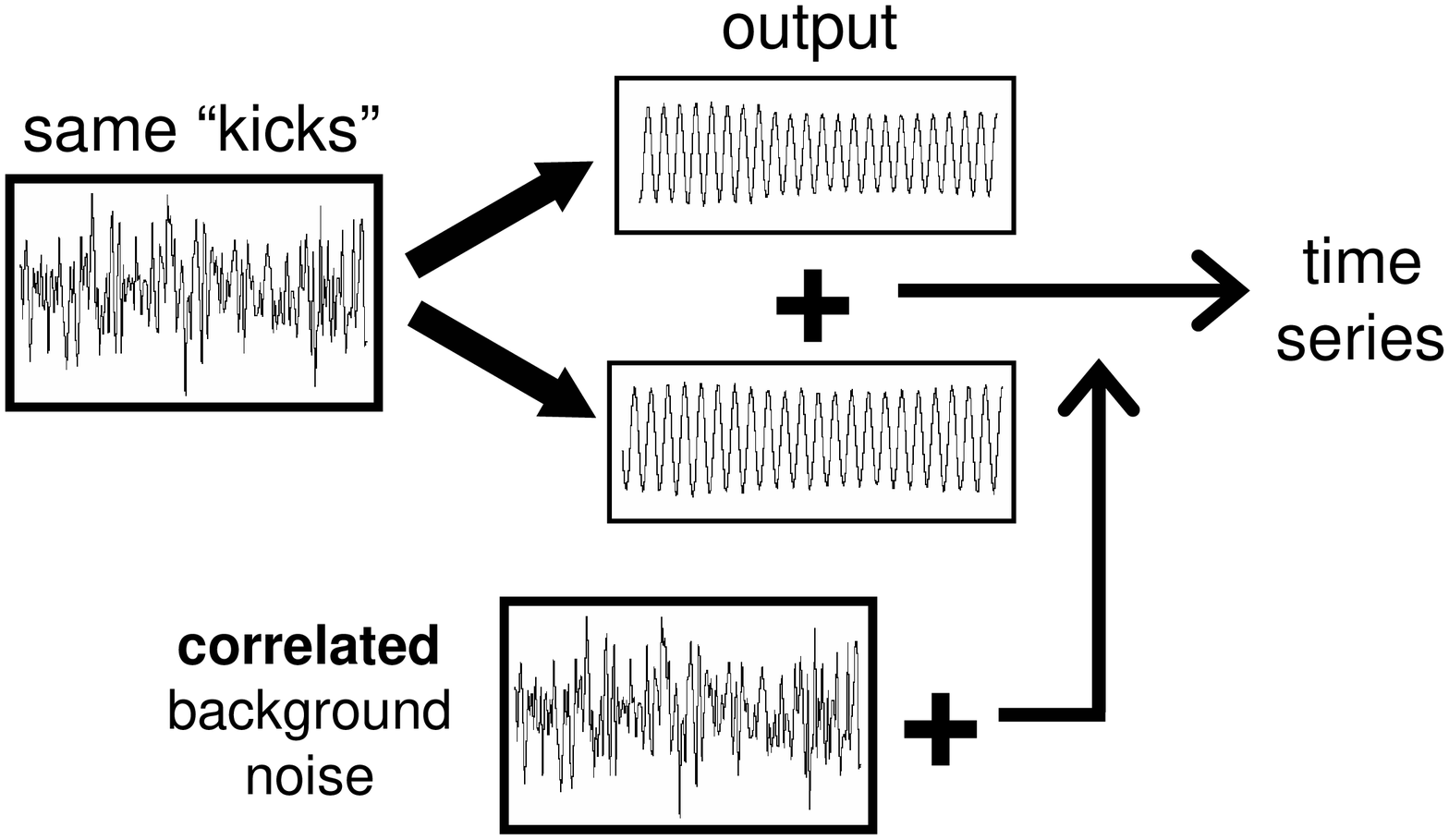}
 \epsfxsize=7.0cm\epsfbox{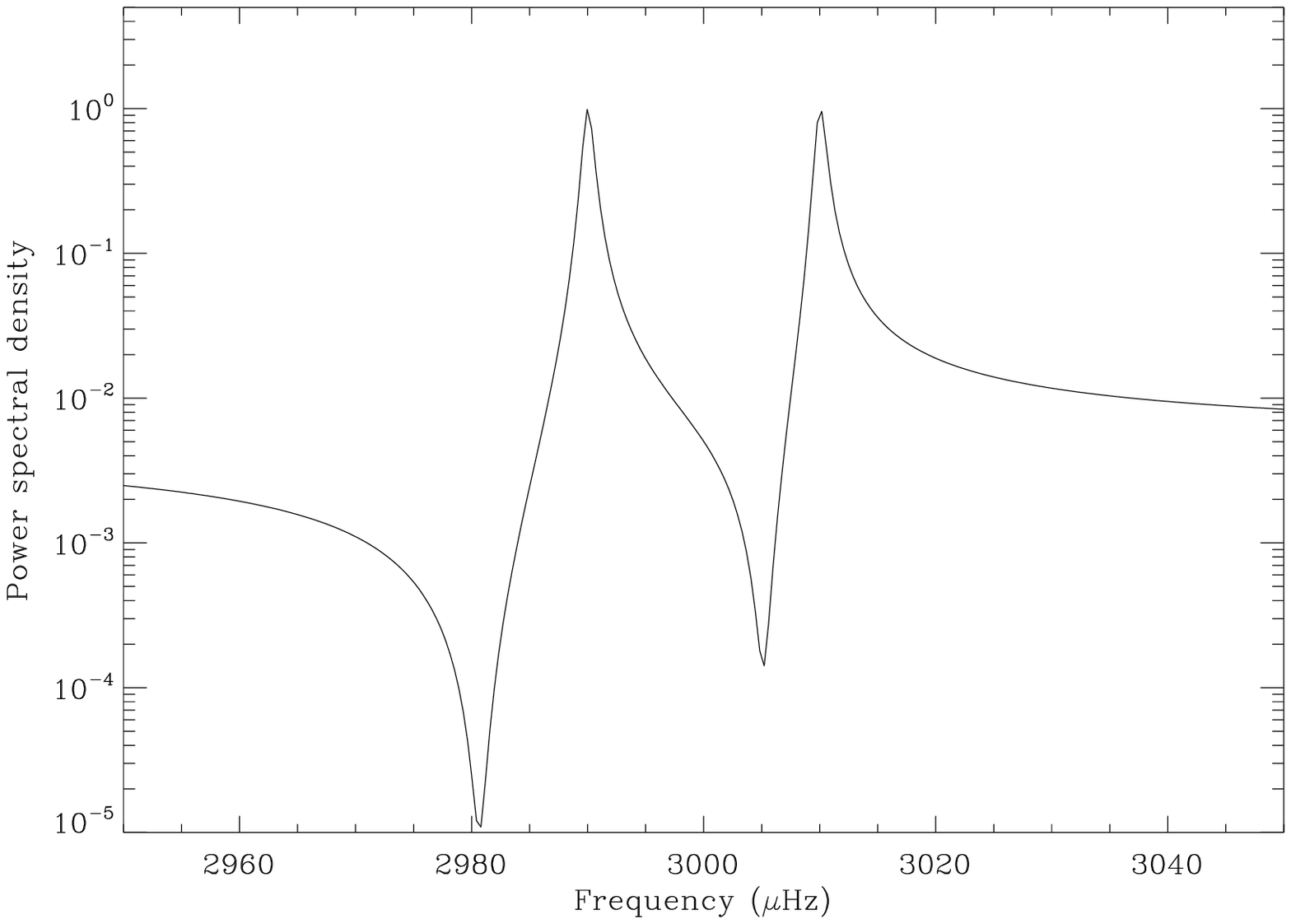}}

 \caption{Case\,\#5: The excitation of two modes is 100\,\%
 correlated, like Case\,\#3; however, here the timeseries also
 includes background noise that is completely correlated with the
 excitation (see Case\,\#1).}

 \label{fig:corr4}
\end{figure*}

Finally, for Case\,\#5 (Figure~\ref{fig:corr4}), we add correlated
background noise. The timeseries of background noise is just the
timeseries of excitation kicks so that, as in Case\,\#1, the
background is 100\,\% correlated with the excitation. We now have two
factors contributing to the observed asymmetry of the peaks. First,
there is the interaction between the two modes, whose excitation is
100\,\% correlated. This contribution was illustrated in
Case\,\#3. Here, we have a second contribution from the correlated
background (like that shown in Case\,\#1), which further distorts the
mode peaks. An analytical description of the power spectral density is
presented in Appendix~\ref{sec:case5}.


 \section{Discussion}
 \label{sec:disc}

Our aim in this paper was to introduce the basic principles and a
mathematical description of correlated p-mode excitation. We used
simplified examples -- involving interactions between two artificial
modes, separated in frequency by just $20\,\rm \mu Hz$ -- to
illustrate the principles.  We saw that when the excitation functions
are correlated in time, the interaction of the wings of the
Lorentzian-like peaks gives rise to peak asymmetry. We also saw that,
provided the frequency separation of the modes is at least several
times the peak linewidths, the output amplitudes of the modes will
\emph{not} be correlated in time; information on the correlations is
instead coded in modifications to the shapes of the mode peaks.

For the Sun, we hypothesize that it is the excitation of overtones of
a given angular degree and azimuthal order that will be correlated in
time.  This follows from the assumption that the excitation function
of a given mode is the same as the component of the solar background
noise that has an identical surface spherical harmonic projection
(over the corresponding range of temporal frequency).  Overtones of
low-degree solar p modes are separated by $\sim 135\,\rm \mu Hz$,
which is somewhat larger than the 20-$\rm \mu Hz$ spacing we used in
the simple two-mode examples in this paper. When there are just two
modes, the impact of the correlated excitation on the shapes of the
resonant peaks will be much smaller when the separation between modes
is $135\,\rm \mu Hz$, as opposed to $20\,\rm \mu Hz$. However, there
are many overtones in the solar spectrum, and so even with the larger
frequency spacing between correlated modes the cumulative effect of
the interactions between all the overtones may be expected to give
rise to significant modifications to the power spectral density.  Peak
asymmetry arising from these correlations would add to the well-known
contribution from the correlated background noise, and would need to
be allowed for explicitly in attempts to make inference on the physics
of the peak asymmetries. We consider the case of the full low-degree
solar p-mode spectrum in Chaplin et al. (in preparation).

\acknowledgements

The authors acknowledge the financial support of the UK Science
Technology and Facilities Council (STFC). WJC thanks T.~Appourchaux
and S.~J.~Jim\'enez-Reyes for useful discussions.

\appendix

 \section{Analytical power spectral density for 2-mode spectra}
 \label{sec:app}

In the following we adopt the naming and symbols conventions from
Section~\ref{sec:recap}. 

 \subsection{Description for Case\,\#3}
 \label{sec:case3}

We begin with a description of the power spectral density of Case\,\#3
(Figure~\ref{fig:corr2}), i.e., a frequency power spectrum comprising
two modes, whose excitation functions are correlated in time, but
there is no background noise. Let the frequency response and
excitation function of the first mode be $L_1(\nu)$ and $E_1(\nu)$,
respectively. The corresponding functions for the second mode are
$L_2(\nu)$ and $E_2(\nu)$, respectively. The observed complex
frequency spectrum is therefore:
 \begin{equation}
 V(\nu)=L_1(\nu)E_1(\nu)+L_2(\nu)E_2(\nu),
 \label{eq:twov}
 \end{equation}
and the observed power spectral density is:
 \[
 P(\nu) = \left<|V(\nu)|^2\right> = 
          \left[ L_1(\nu)E_1(\nu)+L_2(\nu)E_2(\nu) \right] \times
 \]
 \begin{equation}
 ~~~~~~~~~~~~~~~~~~~~~~~~~~~~~~~~~~~~
 \left[ L^*_1(\nu)E^*_1(\nu)+L^*_2(\nu)E^*_2(\nu) \right].
 \label{eq:twop}
 \end{equation}
In what follows we shall drop the explicit dependence of the functions
on $\nu$. Multiplying out the terms in Equation~\ref{eq:twop} above,
we have:
 \[
 P = L_1L_1^*E_1E_1^*+L_1L_2^*E_1E_2^*
 \]
 \begin{equation}
 ~~~~~~~~~+L_1^*L_2E_1^*E_2+L_2L_2^*E_2E_2^*. 
 \label{eq:twop1}
 \end{equation}
Next, we recall from Section~\ref{sec:recap} that the excitation of a
mode is formulated in terms of the real and imaginary parts of a
complex random function. The excitation functions of the two modes may
be written thus:
 \begin{equation}
 E_1 = e_{1r} + ie_{1i},
 \label{eq:e1}
 \end{equation}
and
 \begin{equation}
 E_2 = e_{2r} + ie_{2i}.
 \label{eq:e2}
 \end{equation}
We define a coefficient $\rho$ to describe the correlation of the
excitation of the two modes:
 \begin{equation}
  <e_{1r} e_{2r}> = <e_{1i} e_{2i}> = \rho.
 \label{eq:rho}
 \end{equation}
The coefficient $\rho$ is analogous to the coefficient $\alpha$
(Equation~\ref{eq:alpha}), which was defined in
Section~\ref{sec:recap}: recall that $\alpha$ describes the
correlation of the excitation of a mode with the background noise.
Here, $\rho$ instead fixes the correlation of the excitation of one
mode with another.

The cross term $E_1E_2^*$ in Equation~\ref{eq:twop1} may be written in
terms of the components of Equations~\ref{eq:e1} and~\ref{eq:e2}:
 \[
 E_1E_2^* = [e_{1r} + ie_{1i}] [e_{2r} - ie_{2i}]
 \]
 \[
 ~~~~~~~~~~= e_{1r}e_{2r}+e_{1i}e_{2i}+i[e_{1i}e_{2r}-e_{2i}e_{1r}]
 \]
 \[
 ~~~~~~~~~~= 2\rho + 0 = 2\rho.
 \]
The other cross term, $E_1^*E_2$, is also equal to $2\rho$. We may
also simplify the terms $E_1E_1^*$ and $E_2E_2^*$. Let us take the
first of these:
 \[
 E_1E_1^* = [e_{1r} + ie_{1i}] [e_{1r} - ie_{1i}]
 \]
 \[
 ~~~~~~~~~~= e_{1r}^2+e_{1i}^2 = 2.
 \]
We also have that $E_2E_2^*=2$. The expression for $P$ therefore
simplifies to:
 \begin{equation}
 P = 2L_1L_1^*+2L_2L_2^* + 2\rho[L_1L_2^*+L_1^*L_2].
 \label{eq:twop2}
 \end{equation}
We may then expand out the terms (see Section~\ref{sec:recap}) to
give:
 \begin{equation}
 P = \frac{H_1}{1+x_1^2}+\frac{H_2}{1+x_2^2}+
     2\rho\sqrt{H_1H_2} \frac{1+x_1x_2}{(1+x_1^2)(1+x_2^2)}. 
 \label{eq:twop3}
 \end{equation}
When $\rho=0$, and the excitation of the two modes is independent, we
see that Equation~\ref{eq:twop3} reduces to the sum of two
Lorentzians. This scenario was described as Case\,\#2 in the
paper. When $\rho \ne 0$, and the excitation is correlated, the
observed power spectral density is modified by the third term in
Equation~\ref{eq:twop3}, which contains cross terms between the two
modes. The mode peaks then have asymmetry. When $\rho=1$, this more
complicated scenario corresponds to Case\,\#3.

 \subsection{Description for Case\,\#5}
 \label{sec:case5}

Finally, we consider the description of the power spectral density of
Case\,\#5 (Figure~\ref{fig:corr4}), i.e., a frequency power spectrum
comprising two modes, whose excitation functions are correlated in
time, and there is also correlated background noise. The observed
complex frequency spectrum is now:
 \begin{equation}
 V(\nu)=L_1(\nu)E_1(\nu)+L_2(\nu)E_2(\nu)+N(\nu).
 \label{eq:twovn}
 \end{equation}
The observed power spectral density is therefore:
 \[
 P = 2L_1L_1^*+2L_2L_2^*+2\rho[L_1L_2^*+L_1^*L_2]
 \]
 \[
 ~~~~~~~~+NN^*+N^*[E_1L_1+E_2L_2]+N[L_1^*E_1^*+L_2^*E_2^*].
 \]
The first term on the second line of the equation above is just the
expectation of the noise background described by Equation~\ref{eq:N}
in Section~\ref{sec:recap}, i.e., $NN^*=n$. The other terms in the
second line above depend on the correlation between the excitation and
background noise. Let us take the term $N^*E_1$. It may be written in
terms of the real and imaginary components of $N^*$
(cf. Equation~\ref{eq:N}) and $E_1$ (Equation~\ref{eq:e1})
respectively, i.e.,
 \[
 N^*E_1=\sqrt{\frac{n}{2}}[n_r-in_i] [e_{1r} + ie_{1i}]
 \]
 \[
 ~~~~~~~~~~=\sqrt{\frac{n}{2}}[e_{1r}n_r+e_{1i}n_{i}+i(e_{1i}n_r-e_{1r}n_i)]
 \]
 \[
  ~~~~~~~~~~=2\alpha \sqrt{\frac{n}{2}}+0 = 2\alpha \sqrt{\frac{n}{2}}.
 \]
The terms $N^*E_2$, $NE_1^*$ and $NE_2^*$ simplify to give the same
expression. Putting this all together, the expression for $P$ may
therefore be written as:
 \[
 P = 2L_1L_1^*+2L_2L_2^*+2\rho[L_1L_2^*+L_1^*L_2]
 \]
 \begin{equation}
 ~~~~~~~+n+2\alpha \sqrt{\frac{n}{2}}[L_1+L_2+L_1^*+L_2^*].
 \label{eq:twopn}
 \end{equation}
Equation~\ref{eq:twopn} above comprises two parts. The first part,
which includes the terms on the first line of the equation, is just
the observed power spectral density for two correlated modes, i.e.,
Equations~\ref{eq:twop2} and~\ref{eq:twop3} (Case\,\#3) from
Appendix~\ref{sec:case3}. The second part has a contribution, $n$,
which is just the power spectral density of the background noise, and
further terms which describe the correlation between the modes and the
background noise. These correlated terms modify the observed power
spectral density in a non-trivial manner, and give a contribution to
the observed asymmetry of the mode peaks. 

For completeness, we may take one more step and expand out the terms
in Equation~\ref{eq:twopn} above to give a final expression for the
power spectral density:
 \[
  P = \frac{H_1}{1+x_1^2}\left(1+2\alpha \sqrt{\frac{n}{H_1}}x_1 \right)+
      \frac{H_2}{1+x_2^2}\left(1+2\alpha \sqrt{\frac{n}{H_2}}x_2 \right)
 \]
 \begin{equation}
 + n +2\rho\sqrt{H_1H_2} \frac{1+x_1x_2}{(1+x_1^2)(1+x_2^2)}. 
 \label{eq:twopn1}
 \end{equation}
When $\rho=1$ and $\alpha=1$, Equations~\ref{eq:twopn}
and~\ref{eq:twopn1} describe the power spectral density of Case\,\#5
(Figure~\ref{fig:corr4}).

\end{document}